\documentclass[12pt,preprint]{aastex}

\newcommand{\ai}{{\it ab initio}\ }
\newcommand{\cm}{cm$^{-1}$}

\newcommand{\JCP}{J. Chem. Phys.\ }

\newcommand{\AAA}{A \& A.}
\newcommand{\AAAS}{A \& A S.}
\newcommand{\ApJ}{ApJ}
\newcommand{\ApJS}{ApJS}
\newcommand{\MNRAS}{Mon. Not. R. Astron. Soc.}
\newcommand{\htp}{H$_3^+$}
\newcommand{\MO}{M$_\odot$}
\newcommand{\MIM}{$\mu$m}
\newcommand{\gcc}{g cm$^{-3}$}
\newcommand{\Teff}{T$_{\rm eff}$}

\shorttitle{A new HCN/HNC linelist.}
\shortauthors{Harris et al.}

\begin{document}

\title{The effect of the electron donor \htp\  on the pre-main and Main Sequence evolution of low mass zero metallicity stars.}

\author{G. J. Harris, A. E. Lynas-Gray\altaffilmark{1}, S. Miller and J. Tennyson\altaffilmark{2}.}
\affil{Department of Physics and Astronomy, University College London, London, WC1E 6BT, UK.}

%\email{mail to: j.tennyson@ucl.ac.uk}

\altaffiltext{1}{Permanent address: Department of Physics, University of Oxford, Keble Road, Oxford OX1 3RH, UK.}        
\altaffiltext{2}{Corresponding Author:  j.tennyson@ucl.ac.uk}
                                         
\begin{abstract}

 \htp\  has been shown (Lenzuni et al. 1991, \ApJS, 76, 759) to be the dominant positive ion, in a zero metallicity gas at low temperature and intermediate to high density.
It therefore affects both the number of free electrons and the opacity of the gas. The most recent \htp\  partition function (Neale and Tennyson, 1995, \ApJ, 454, L169)
is an order of magnitude larger at 4000 K than all previous partition functions, implying that \htp\  is a more important electron donor than previously thought.

Here we present new Rosseland mean opacities for a hydrogen-helium gas of 1000 $\le$ T(K) $\le$ 9000  and $-14 \le \log_{10}(\rho\  [{\rm g~cm^{-3}}]) \le-2$. In the calculation of these opacities we have made use of the latest collision induced absorption data as well as the most recent \htp\  partition function and line opacity data. It is shown that these updated and new sources of opacity give rise to a Rosseland mean opacity for a hydrogen-helium gas which is in general greater than that calculated in earlier works.

The new opacity data are then used to model the evolution of low mass 0.15-0.8 \MO\  zero metallicity stars, from pre-Main Sequence collapse to Main Sequence turn-off. To investigate the effect of \htp\  on the evolution of low mass zero metallicity stars, we repeat our calculations neglecting \htp\  as a source of electrons and line opacity.
We find that \htp\  can have an effect on the structure and evolution of stars of mass $\sim$0.5 \MO\  or less. A gray atmosphere is used for the calculation, which is sufficient to demonstrate that \htp\  affects the evolution of very low mass stars to a greater degree than previously believed.

\end{abstract}

\section{Introduction.}

It has long been speculated that zero metallicity, Population III stars exist, but as yet none of these stars have been identified. At the current epoch low mass stars are the most numerous stars in the Galaxy, however the first stars are often thought to have formed with exclusively high masses. For example \citet{Tohline} estimated the minimum mass of a Population III star to be a massive 1500 \MO, more recently \citet{Araujo} estimated the minimum mass to be 50 \MO. On the other hand \citet{Kashlinsky} proposed a method by which stars of mass as low as 0.2 \MO\  could be formed from primordial material. Recently using hydrodynamical simulations \citet{Nakamura} suggested a bi-modal Population III initial mass function with a minimum mass of between 1 and 2 \MO.
The discovery of the extremely metal deficient star HE0107-5240 \citep{Christ} which has an estimated mass of 0.8 \MO\  and [Fe/H]=-5.3 has added credence to the possibility that low mass primordial stars formed. All Population III stars with mass less than about 0.75 \MO\  should still be on the Main Sequence and thus could, if ever observed, provide an estimate of the age of the Universe. 

The accuracy of stellar formation and evolution calculations is, among other things, dependent upon available opacity data. In the absence of metals the only contributors to the opacity of the atmospheres Population III stars are the various hydrogen and helium species and ions, and electrons. In a zero metallicity gas at densities common to stellar atmospheres it is H$^-$ that dominates opacity at temperatures between about 3500 K and 7000 K. Below about 3500 K it is collision induced absorption (CIA) by  H$_2$-He and H$_2$-H$_2$ that dominates the opacity. Many authors \citep{Lenzuni,Alexander,Saumon,Rohr2001,Allard}  have commented upon how H$_3^+$ can act as an electron donor in low metallicity stars, thereby increasing the abundance of H$^-$ and hence the opacity. However none have yet investigated the effect of \htp\  rotation-vibration line absorption on a hydrogen-helium gas opacity.

The zero metallicity opacity calculations of \citet{Lenzuni} used the \htp\  partition function of \citet{Patch}, whereas the opacity calculations of \citet{Alexander}, \citet{Saumon} and \citet{Bergeron} used the partition functions of \citet{Chandra}. The partition functions of \citet{Patch} and \citet{Chandra} are an order of magnitude smaller at T$\sim$4000 K, than the latest most accurate \htp\  partition function of \citet{Neale95}. The \citet{Neale95} \htp\  partition function was computed by direct summation of 200~000 rotation-vibration energy levels.
This most recent \htp\  partition function has been used extensively in the modeling of cool stellar, white and brown dwarf atmospheres \citep{Bergeron97,Hauschildt99,Saumon99,Rohr2001,Allard}. The cool stellar and brown dwarf atmospheres of \citet{Hauschildt99} have subsequently been incorporated in evolution models of brown dwarfs and low mass stars by \citet{Chabrier97} and \citet{Baraffe97,Baraffe98,Baraffe02}. In particular \citet{Baraffe97,Baraffe98} studied very low mass, low metallicity stars of [M/H]$\geq-$2.0 and \citet{Saumon} studied zero metallicity very low mass stars with the \citet{Chandra} partition function.
However neither the \htp\  partition function of \citet{Neale95} or any \htp\  linelist has been used in evolution calculations of low mass, zero or very low metallicity stars. Recent very low and zero metallicity low mass stellar evolution calculations  by \citet{VandenBerg}, \citet{Marigo}, and \citet{Siess} use the low temperature (T$<10~000$ K) \citet{Alexander} opacities. Whereas \citet{Richard} use low temperature opacities based on Kurucz data, see \citet{Turcotte} and \citet{Proffitt}. The minimum mass to which these very low metallicity evolution calculations extend down to is 0.5 \MO. As we show below, zero metallicity stars of mass less than about 0.5 \MO\  are affected by \htp\  acting as an electron donor.

In this paper we present new low temperature Rosseland mean opacities for a zero metallicity gas in LTE. These opacities have been calculated both with and without the \citet{Neale95} partition function and the \citet{Neale96} \htp\  linelist. We then incorporate these new opacity data within the CESAM stellar evolution code \citep{Morel} and examine the effect of \htp\  on the pre-main and Main Sequence evolution of zero metallicity stars.

\section{Low temperature opacity.}

For compatibility with the CESAM stellar evolution code \citep{Morel}, the Rosseland mean opacity must be determined from the state variables temperature (T), mass density ($\rho$) and hydrogen mass fraction (X). The opacity of a gas is dependent upon its chemical composition, so the abundances of all the species within the gas or plasma must first be determined. For this purpose thermodynamic and chemical equilibrium are assumed, which for low densities allows the uses of the Saha equation to determine the number densities of individual species.
 By combining these number densities with the cross sections of the relevant scattering or absorption process we can obtain the total opacity for the process in the gas.
Rosseland mean opacities are then computed as below.

\subsection{The equation of state.}
\label{subsec:EoS}

The species that are considered in this work are listed in table \ref{tab:partfunc} together with the source of the rotational-vibrational partition function if relevant.
We use an iterative approach to solve for chemical equilibrium using the Saha equations for each chemical species. 

For the initial conditions of our calculation a neutral H$_2$ and He gas is used, the number densities of all other species are set to zero. The initial number densities of H$_2$ and He are determined from temperature, $\rho$ and X. For most of the values of T and $\rho$ considered in this work H$_2$, H or He are the most abundant species. So we first determine the dominant hydrogen species by calculating a first order number density of H from H$_2$ using the Saha equation:
\begin{equation}
\frac{n({\rm H_2})}{n({\rm H})^2}=\frac{Q_T({\rm H_2})}{Q_T({\rm H})^2}\exp \left(\frac{\chi({H_2})}{kT}\right)
\label{eq:H2H}
\end{equation}
where $n(x)$ is the number density of the species $x$, $Q_T(x)$ is the total partition function of species $x$, and $\chi({\rm H_2})$ is the dissociation energy of H$_2$. The total partition function, $Q_T(x)$, which for H$_2$ is the product of translational, rotational-vibrational and electronic partition functions. In this work we neglect electronic states higher than the ground state and set the electronic partition function equal to the degeneracy of the ground electronic state. 

Equation \ref{eq:H2H} is solved iteratively using a Newton-Raphson technique and forcing the conservation of H nuclei.
Using the Saha ionization equation we then determine the first order number densities of H$^+$, H$^-$, He$^+$, H$_2^+$, H$_2^-$, HeH$^+$ and $e^-$. Again an iterative Newton-Raphson technique is employed, taking care to conserve charge and hydrogen and helium nuclei.

The determination of the number density of \htp\  is slightly more difficult. The usual formation reaction of \htp\  in the interstellar medium is:
\begin{equation}
\rm H_2^++H_2 \rightleftharpoons H_3^++H
\end{equation}
see for example \citet{Herbst}.
In chemical equilibrium the number density of H$_2^+$ is often many orders of magnitude smaller than that of H$_3^+$ which in turn is many orders of magnitude smaller than the number densities of H$_2$ and H. This can result in problems with machine accuracy. Instead we estimate the number density of \htp\  from H$_2$ or H, depending upon whether H$_2$ or H is more abundant. For instance if H$_2$ is dominant then we use the hypothetical reaction: $\rm 2H_2 \rightleftharpoons H_3^++H+e^-$. This reaction has the Saha equation:
\begin{equation}
\frac{n({\rm H_2})^2}{n({\rm H_3^+})n({\rm H})n(e^-)}=\frac{[Q_T({\rm H_2})]^2}{Q_T({\rm H_3^+})Q_T({\rm H}) Q_T(e^-)}\exp \left(\frac{\chi}{kT}\right)
\end{equation}
where $\chi$ is the difference in energy between products and reactants, this is equal to the dissociation energy of H$_2$ (4.52 eV) plus the ionization potential of H (13.60 eV) minus the proton affinity of H$_2$ (4.38 eV), so $\chi$=13.74 eV.
This approach increases the stability of the calculation of the number density of \htp. It is valid because, by definition, in chemical equilibrium each species is in equilibrium with every other species. 

Once the first estimate for the number density of \htp\  has been obtained, the process is reiterated, using the new number densities, until converged values for the number density of each species are obtained.

\begin{figure}
\epsscale{0.9}
\plotone{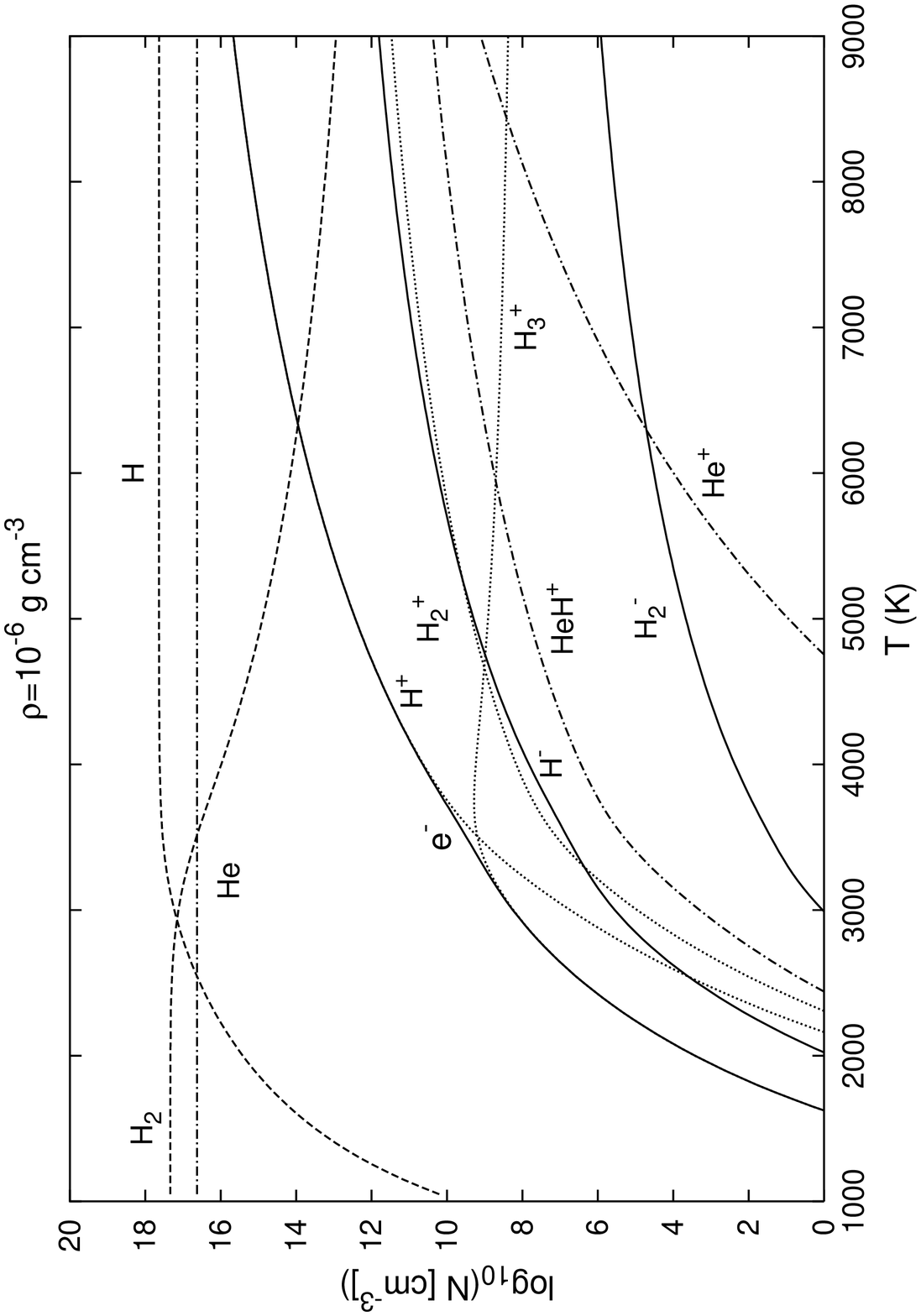}
\caption{The number densities of the species considered in the calculation as a function of temperature, at $\rho=10^{-6}$ \gcc. The solid curves represent negatively charged species, the dotted curves represent positively charged hydrogen species, the dashed curves represent neutral hydrogen species and the dot-dashed curves represent helium containing species.}
\label{fig:N_Vs_T_all}
\end{figure}

\begin{figure}
\epsscale{0.9}
\plotone{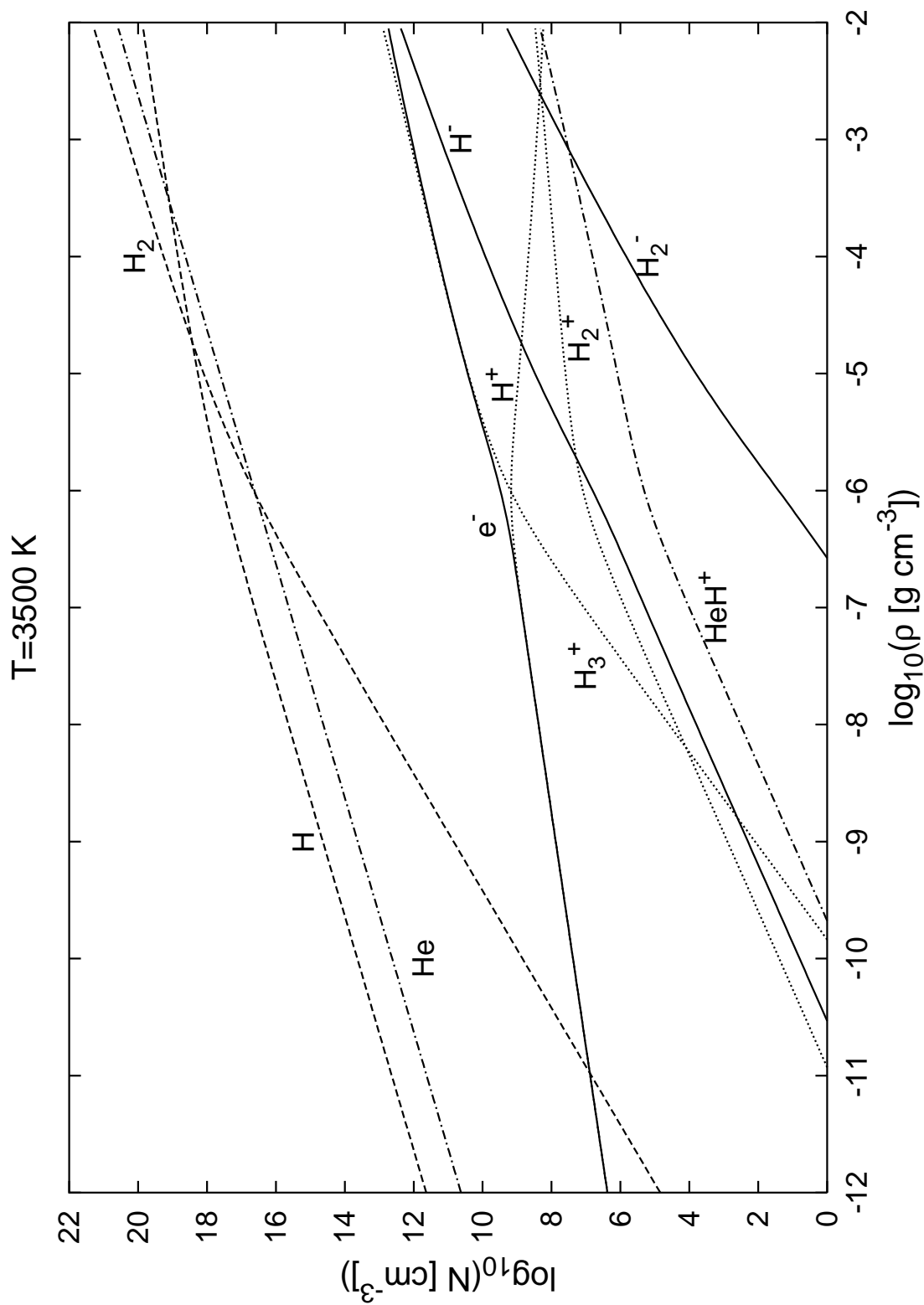}
\caption{The number densities of the species considered in the calculation as a function of density, at T=3500 K. The solid curves represent negatively charged species, the dotted curves represent positively charged hydrogen species, the dashed curves represent neutral hydrogen species and the dot-dashed curves represent helium containing species.}
\label{fig:N_Vs_ro_all}
\end{figure}

\begin{figure}
\epsscale{0.9}
\plotone{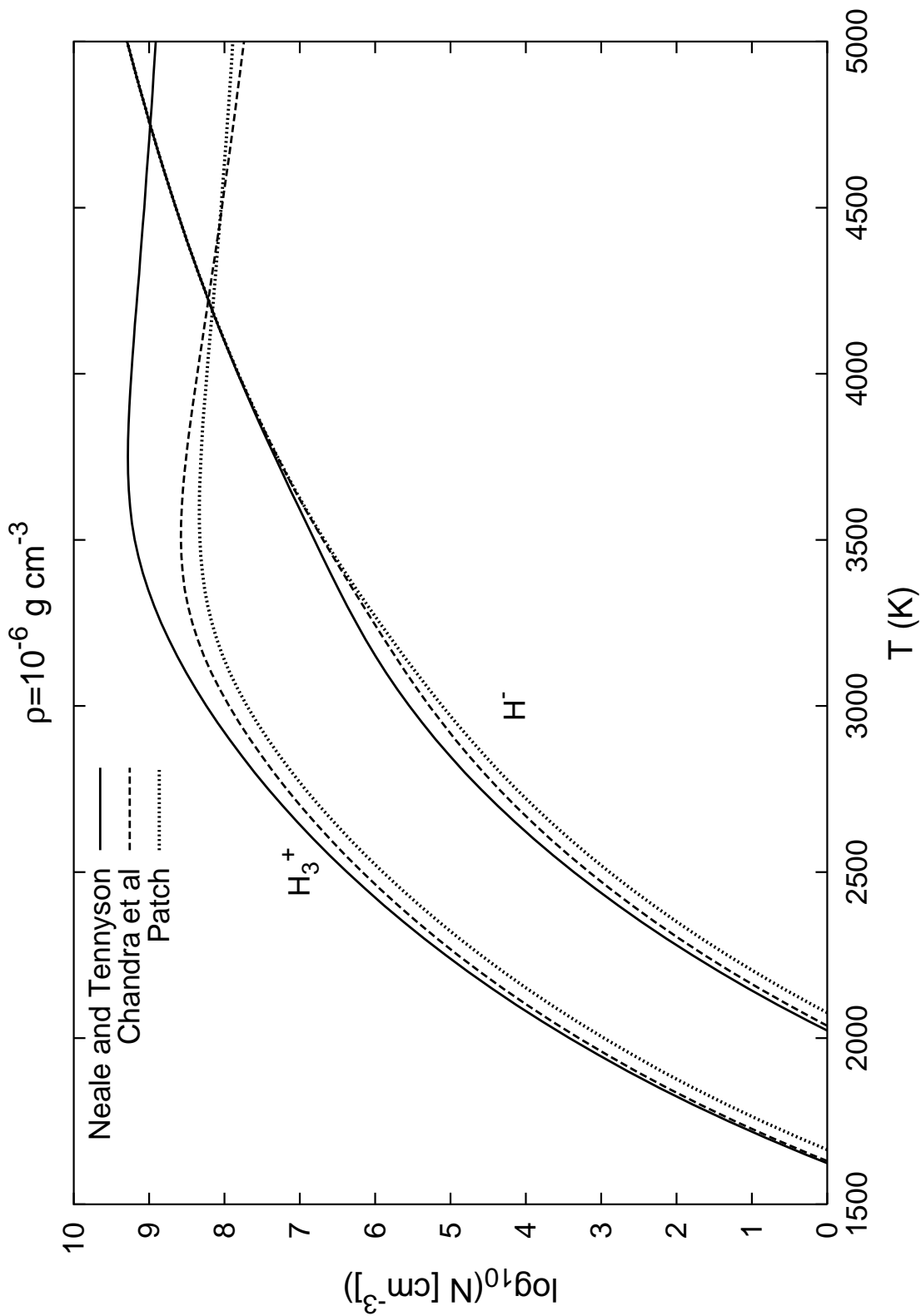}
\caption{The number densities of \htp\ and H$^-$ as a function of temperature, at $\rho=$10$^{-6}$ g cm$^{-3}$. The three curves were calculated with the \htp\  partition functions of \citet{Neale95} solid curves, \citet{Chandra} dashed curves and \citet{Patch} dotted curves.}
\label{fig:N_Vs_T_h3p}
\end{figure}

Figure \ref{fig:N_Vs_T_all} shows an illustration of the temperature dependences of the number densities of each species at $\rho=10^{-6}$ \gcc\  and X=0.72. Importantly at $\rho=10^{-6}$ \gcc\  \htp\  is the dominant positive ion below 3500 K, so will have a strong effect on the abundances of ions and electrons below this temperature. 
Figure \ref{fig:N_Vs_ro_all} shows an illustration of the mass density dependences of the number densities of each species at T=3500 and X=0.72. Here \htp\  becomes the dominant ion above $\rho \sim 2 \times 10^{-6}$. Figure \ref{fig:N_Vs_T_h3p} shows the number density of H$^-$ and \htp\  calculated using the \htp\  rotational-vibrational partition functions of \citet{Neale95}, \citet{Chandra} and \citet{Patch} as a function of temperature with $\rho=10^{-6}$ and X=0.72. Clearly the number densities of \htp\  and H$^-$ are strongly dependent upon which partition function is used, indeed above 4000 K use of the \citet{Neale95} partition function yields an order of magnitude more \htp\  than the use of other partition functions. However at this mass density H$^+$ becomes the dominant ion above 3700 K and the effect of \htp\  on H$^-$ quickly diminishes with increasing temperature. As H$^-$ is an important source of opacity, the opacity is sensitive to which partition function is used in the calculation.

The limits of the partition functions used in this work effectively determine the temperature limits for which this equation of state is reliable. For all but \htp\  the partition functions extend up to at least 9000 K, and down to 1000 K. For \htp\  the \citet{Neale95} partition function extends to 8000 K. At the densities of interest to us \htp\  is relatively unimportant between 8000 and 9000 K. Therefore our equation of state is reliable between 1000 to 9000 K. The main short coming of the Saha equation is that it does not account for pressure ionization which occurs at very high densities. This imposes an upper density limit above which our equation of state is not valid. \citet{Mihalas} show that for T less than 10~000 K, the upper density for which the Saha equation is valid is around $\log \rho=-2$, we take this as the upper density limit for our equation of state.

\subsubsection{The effect of metals on the number density of \htp.}

\htp\  is readily destroyed in a dissociative recombination reaction with electrons. Hence high abundances of electrons act to reduce the abundance of \htp. Many species of metals have an ionization energy that is significantly lower than that of hydrogen, so that, in a gas containing metals, it is the metals that govern the abundance of free electrons. In this way metals have a strong effect on the abundance of \htp. As all known stars contain some metals we briefly investigate the effect that these metals have upon \htp.

The equation of state is dependent upon the relative abundance of each metal. As the stars in which \htp\  is an important electron donor are likely to be metal poor 
we use the metal abundance mix found in the very metal poor star HE0107-5240, discovered by \citet{Christ}. Where the abundances of these metals are unmeasured for HE0107-5240 we use the metal abundances based upon a theoretical Population III supernova yield calculated by \citet{Umeda}. 
In our equation of state we include what are likely to be the ten most abundant metals in HE0107-5240: C, N, O, Ne, Na, Mg, Si, S, Ca and Fe. We included the first ionization states, and the anions of the metals within our equation of state.

\begin{figure}
\epsscale{0.9}
\plotone{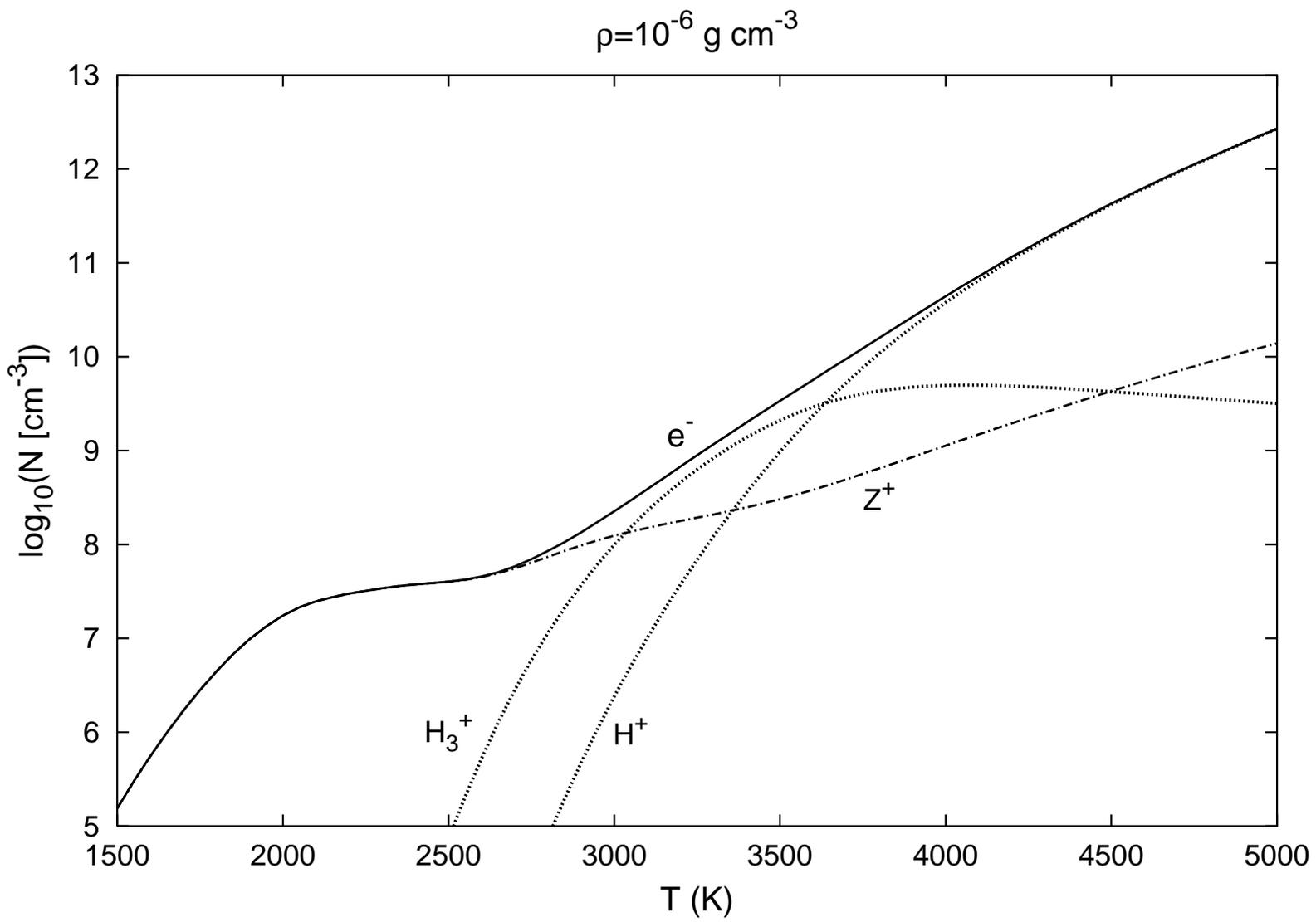}
\caption{The number densities of \htp, H$^+$, e$^-$ and ionized metals (Z$^+$) as a function of temperature, at $\rho=$10$^{-6}$ g cm$^{-3}$.}
\label{fig:metals1}
\end{figure}

\begin{figure}
\epsscale{0.9}
\plotone{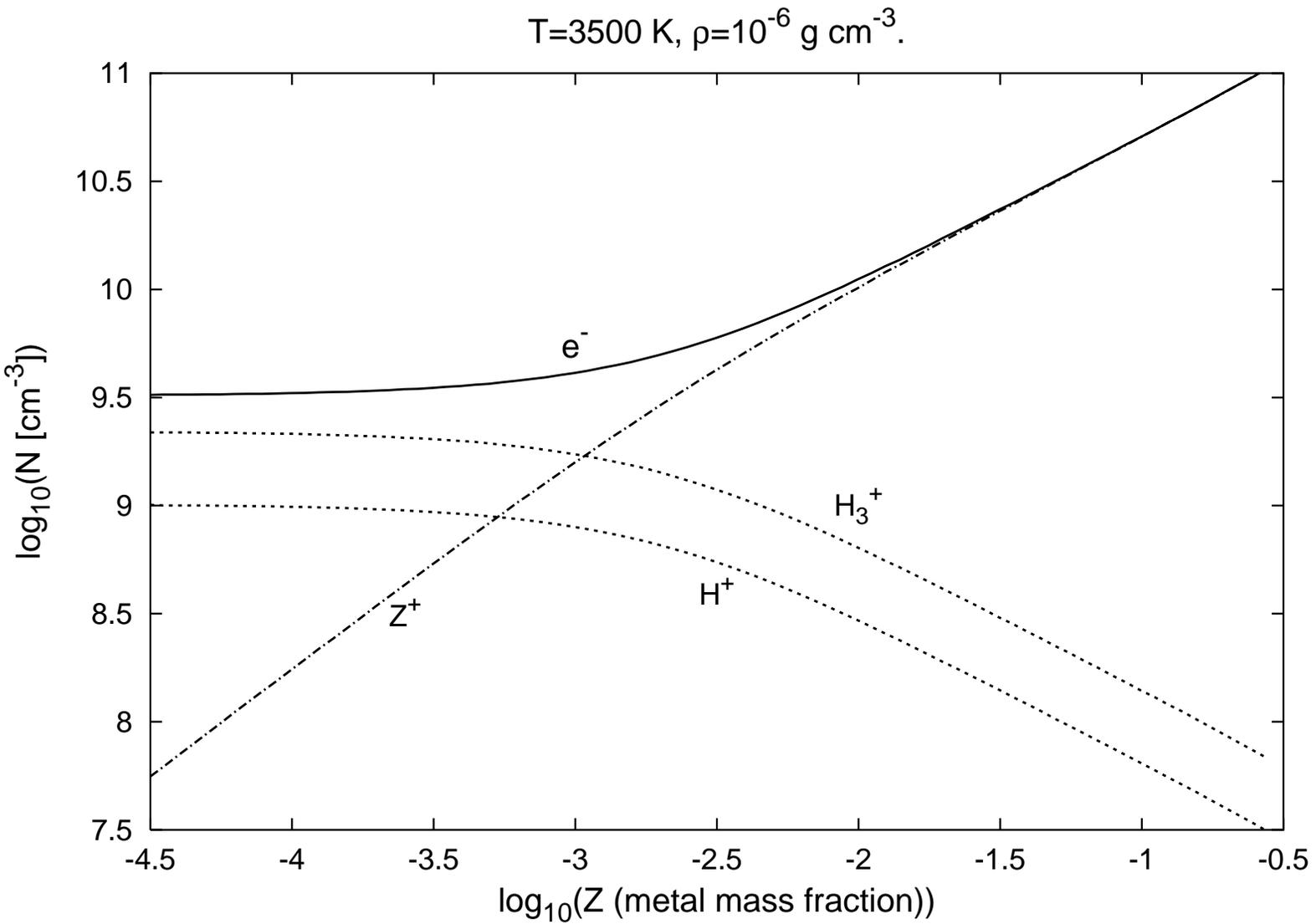}
\caption{The number densities of \htp, H$^+$, e$^-$ and ionized metals (Z$^+$) as a function of metal mass fraction, at T$=$3500 K and $\rho=$10$^{-6}$ g cm$^{-3}$.}
\label{fig:metals2}
\end{figure}

Figure \ref{fig:metals1} shows the number densities of \htp, H$^+$, e$^-$ and the positive metal ions (Z$^+$) as function of temperature, at a density of 10$^{-6}$ \gcc. From this figure, \htp\  is the dominant positive ion in a gas between temperatures of 3000 and 3600 K with a elemental abundances similar to HE0107-5240. 
Figure \ref{fig:metals2} shows the number densities of \htp, H$^+$, e$^-$ and the positive metal ions (Z$^+$), as function of metal mass fraction. We have used the same metal mix as for figure \ref{fig:metals1} with a constant hydrogen mass fraction of X$=$0.72 and a density of 10$^{-6}$ \gcc. The total number density of the positive metal ions only exceeds the number density of \htp\  for metal mass fractions above $10^{-3}$. It should be noted that the metal mix found in HE0107-5240 has a very high carbon to iron ratio relative to solar ([C$/$Fe]$=4.0$), but has a comparatively low abundance of more readily ionized elements such as Na ([Na$/$Fe]$=0.8$) and Fe. Hence this metal mix will result in fewer free electrons at low temperatures than a solar metal mix. For a solar metal mix \citep{Grevesse} with the same temperature and density, we find that the metals are the main electron donors for metal mass fractions above about 10$^{-6}$.
The relative importance of \htp\  as an electron donor is not only dependent upon metallicity, but also the relative abundances of the metals within the mix.

Density also has a role in determining the dominant positive ion, high densities favor \htp\  low densities favor the metal ions. Hence \htp\  is more likely to have an effect on very metal poor low mass dwarf stars rather than low mass giants.

\subsection{The opacity.}

The frequency dependent continuous opacity can now be calculated using the number densities of the various species.
In this work the set of subroutines developed by \citet{opa_codes}
 were used to calculate the continuous opacity contributions from H~I, H$^-$, He$^-$, He~I, He~II bound-free and free-free, H$_2^-$ free-free, Rayleigh scattering of He~I, H$_2$ and H~I and Thompson scattering by e$^-$. 
We have also included the \citet{Lebedev}  H$_2^+$ free-free and bound-free opacity data. 
The method used by \citet{opa_codes} for each absorption or scattering process is summarized in table \ref{tab:opasour} along with all the other opacity sources. We have not included any photodissociation processes for HeH$^+$ as these are considered unimportant or important only at extreme ultra violet wavelengths \citep{Roberge}. We have also included H~I line absorption by using the STARK subroutine from the ATLAS12 \citep{Kurucz1993} program.
H~I line absorption is important for temperatures greater than $\sim$7500 K.

In this work we make use of the most recent CIA data, which are in a tabular form. We use the H$_2$-He CIA data of \citet{CIA_H2He}, which covers the temperature range 1000 to 7000 K and the wavenumber range 25 to 20~000 \cm. The H$_2$-H$_2$ data of \citet{CIA_H2H2} which covers the temperature range 1000 to 7000 K and the wavenumber range 20 to 20~000 \cm. The H-He data of \citet{CIA_HHe} which covers the temperature range 1500 to 10~000 K and the wavenumber range 50 to 11~000 \cm. No CIA data for H-He collisions were available to \citet{Lenzuni} and \citet{Alexander}, but these data have been used by \citet{Rohr} in the calculation of synthetic white dwarf spectra.
Our calculations indicate that at low hydrogen mass fractions and $\log \rho \sim -8$ and $\log T \sim 3.5$ H-He CIA contributes significantly to the Rosseland mean opacity and is the dominant source of continuous opacity between wavelengths of $\sim$3.3 to $\sim$13 \MIM. We have neglected H-H$_2$ CIA, however, since it has only a small effect on opacity in a narrow temperature range when H$_2$ and H are of near equal abundances. At low hydrogen mass fractions, low temperatures and moderate to high densities He-He CIA may have an effect on opacity. We have not found any He-He opacity data, so have not been able to include this source of opacity in our calculations.

For the purposes of our calculation the CIA opacity tables of \citet{CIA_H2He,CIA_H2H2,CIA_HHe} are interpolated using a bicubic spline. Frequencies at which CIA is an important source of opacities are covered by the frequency ranges of the data tables. Beyond the upper temperature limit of the CIA data tables many opacity sources exceed, by several orders of magnitude, the opacity from CIA. However CIA from H$_2$-He and H$_2$-H$_2$ collisions remains important for some temperatures below 1000 K. So the calculation of the total opacity is limited only by the lower temperature limits of the H$_2$-He and H$_2$-H$_2$ CIA data tables, which is 1000 K.

To investigate the effect of direct absorption by \htp\  the \ai\  rotational-vibrational linelist of \citet{Neale96} was used. This linelist contains over 3$\times10^6$ lines between 0 and 15~000 \cm\  all of which were used.
 A thermal Doppler line profile was used to distribute the intensity of each line, this approximation to the true line profile is valid as all the \htp\  lines are weak such that the extended line wings contribute little to the overall opacity. \htp\  was found to contribute up to 15\% of the Rosseland mean opacity via line absorption. This is far smaller than the indirect effect \htp\  has on opacity through electron donation, which can increase the Rosseland mean opacity by over a factor of 3. However it implies that \htp\  lines may well be observable in cool Population III stellar spectra. We have identified a mistake in figures 1 and 2 of \citet{Neale96}. An incorrect degeneracy factor of 2 instead of 8/3 was used to calculate the absorption by hotter lines in these figures, which resulted in the quoted absorption and emission being reduced by up to 25\%. The mistake affects only figures 1 and 2 and has no bearing on the Einstein A coefficents of the \citet{Neale96} linelist.

The ion HeH$^+$ has a strong electric dipole moment and so will have a strong infrared spectrum. We have not been able to find an extensive set of HeH$^+$ line intensities, so we cannot include HeH$^+$ line absorption into our opacity calculation. The line opacity of \htp\  only contributes significantly to the Rosseland mean opacity when the opacity of the gas is low and the number density of \htp\  is high. The number densities of HeH$^+$ exceed \htp\  only at high temperatures, when the Rosseland mean opacity is high.
We would expect the total line absorption intensities of HeH$^+$ per molecule to be weaker than that of \htp. So we believe that HeH$^+$ will have little effect on the opacity, but this needs to be verified.

\begin{figure}
\epsscale{0.9}
\plotone{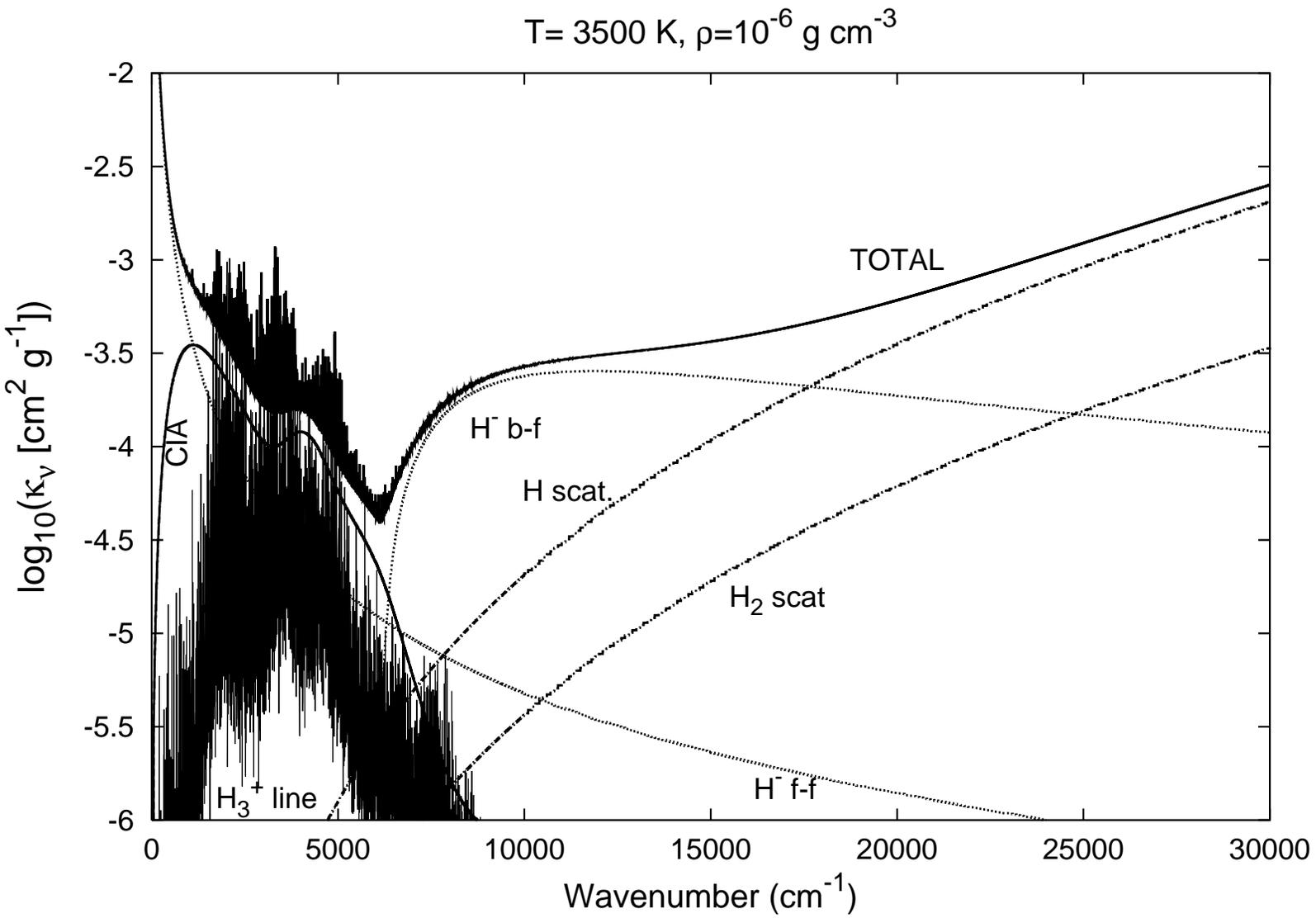}
\caption{The dominant sources of monochromatic opacity as a function of wavenumber at 3500 K and $\rho=$10$^{-6}$ g cm$^{-3}$.}
\label{fig:mono}
\end{figure}

Figure \ref{fig:mono} shows the frequency dependent opacity of the dominant contributors to the total continuous opacity at T=3500 K, $\rho=10^{-6}$ g cm$^{-3}$ and X=0.72. The total opacity is calculated by summing all the contributors to the opacity. This figure illustrates the strength of CIA, H$^-$ and \htp\  line absorption, hence  the importance that these sources of opacity are correctly accounted for. The \htp\  line absorption has a clear effect on the frequency dependent opacity between 1500 and 8000 \cm. This indicates that line absorption may well be observable in certain cool dense metal free stars, brown and possibly cool white dwarfs. 

The Rosseland mean opacity is defined by:
\begin{equation}
\frac{1}{\kappa_R}=\frac{\int_{0}^{\infty} \frac{1}{\kappa_\nu} \frac{\partial B_\nu}{\partial T} d\nu}{\int_{0}^{\infty} \frac{\partial B_\nu}{\partial T} d\nu}
\end{equation}
Where $B_\nu$ is the Planck function, $\kappa_R$ is the Rosseland mean opacity and $\kappa_\nu$ is the monochromatic opacity.
The denominator is evaluated analytically, but the numerator must be evaluated numerically. We use 30~000 integration points, which is sufficient to account for \htp\  line absorption and to converge the integration. The partition functions we have used to calculate the number densities of the various species limit us to temperatures above 1000 K. We therefore only give opacity data for temperatures above 1000 K. This minimum temperature is sufficient for our stellar models.

\begin{figure}
\epsscale{0.9}
\plotone{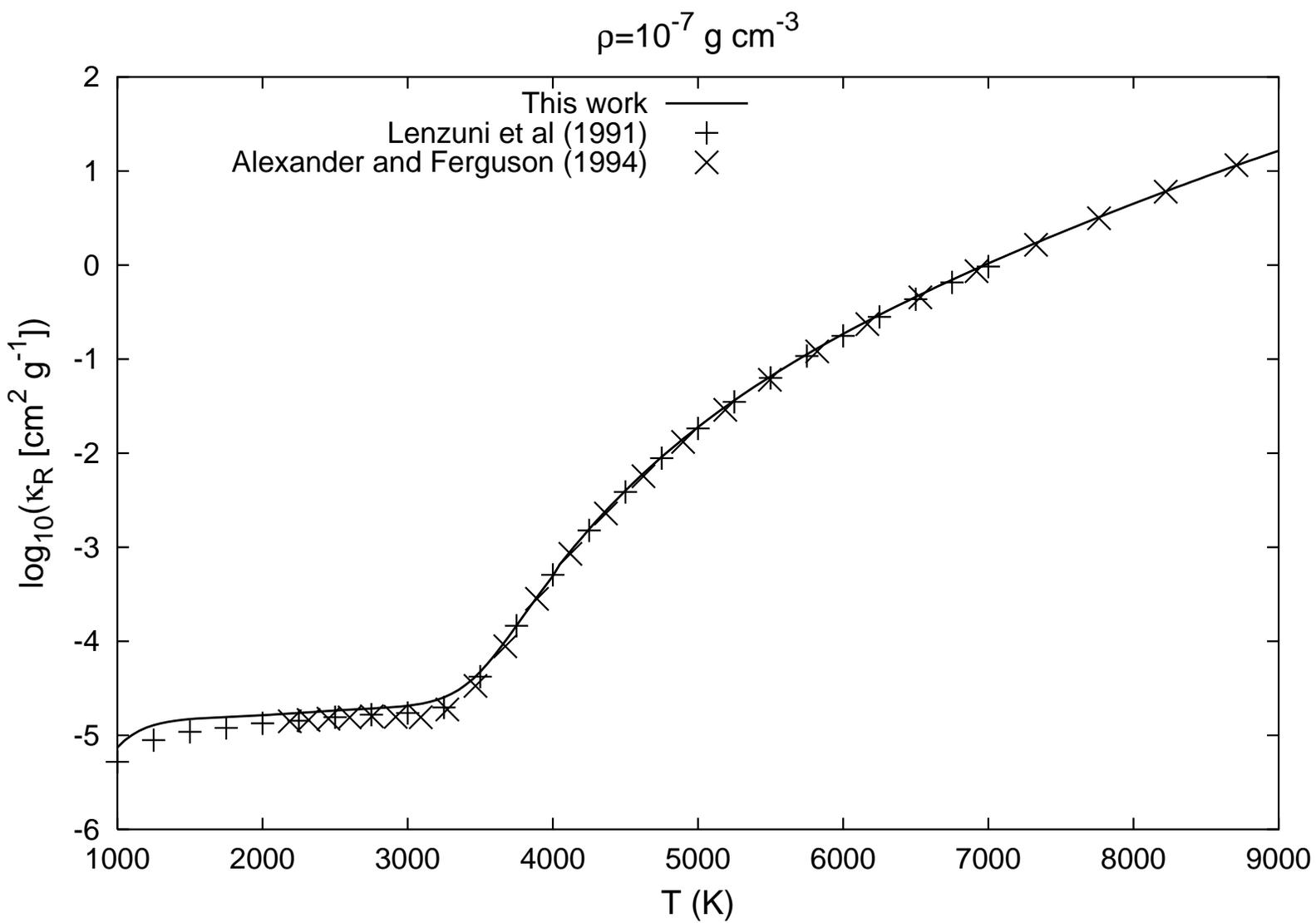}
\caption{A comparison of available Rosseland mean opacity for a hydrogen-helium gas at $\rho=$10$^{-7}$ g cm$^{-3}$.}
\label{fig:ross_prev}
\end{figure}

Figure \ref{fig:ross_prev} show the Rosseland mean opacity calculated in this work along with those of \citet{Lenzuni} and \citet{Alexander} at $\rho=10^{-7}$ and X=0.72. The \citet{Alexander} data were interpolated using a bi-cubic spline.
 A density of $\rho=10^{-7}$ is too low for the effect of \htp\  on the number density of H$^-$ to be evident. However at this density \htp\  line opacity is strong enough to contribute 13 \%  of the Rosseland mean opacity at 3250 K.
The Rosseland mean opacity of this work for T$<3500$ K is significantly higher than those of \citet{Lenzuni} and \citet{Alexander}. At these temperatures the opacity is dominated by CIA and a little \htp\  line opacity. The older works of  \citet{Lenzuni} and \citet{Alexander} do not include \htp\  line opacity and use older CIA data which is weaker than the more recent CIA data, see \citet{CIA_H2H2} and \citet{CIA_H2He}. So a greater opacity at these temperatures is to be expected.

\begin{figure}
\epsscale{0.9}
\plotone{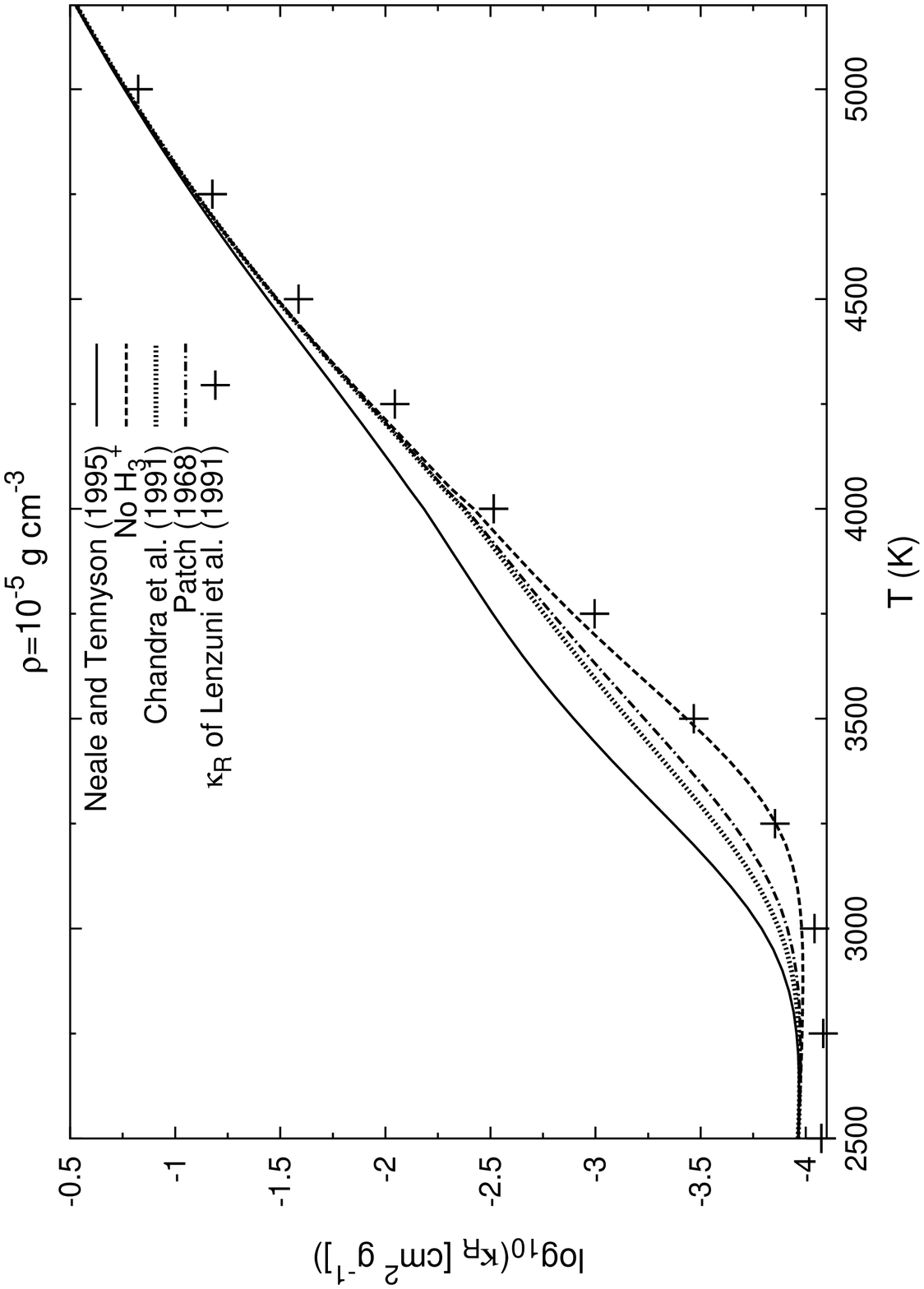}
\caption{A comparison of Rosseland mean opacity at $\rho=10^{-5}$ calculated with the partition functions of \citet{Neale95}, \citet{Chandra}, \citet{Patch} and without \htp. The Rosseland opacity of \citet{Lenzuni} is also plotted.}
\label{fig:ross_Q}
\end{figure}

Figure \ref{fig:ross_Q} shows a plot of the Roseland mean opacity at $\rho=10^{-5}$ \gcc\  and X=0.72. The opacities were calculated with the \htp\  partition functions of \citet{Patch}, \citet{Neale95}, \citet{Chandra} and also without accounting for \htp. The Rosseland mean opacity of \citet{Lenzuni} is also plotted, but that of \citet{Alexander} does not extend to these high densities. Again the difference between the opacity of \citet{Lenzuni} and this work are evident, at these temperatures collision induced absorption is important for T$<4500$ K and accounts for part of the difference. \htp\  line absorption is weak making up $\sim$1.1\% of the Rosseland mean opacity at 4000 K.
But the most important factor is clearly attributable to the effect of \htp\  on the opacity. The partition functions of \citet{Patch} and \citet{Chandra} lead to an under prediction of the abundance of H$^-$  and hence the opacity. It would appear that \htp\  has an effect on the opacity several times stronger, and to about 500 K hotter, than predicted by the partition functions of \citet{Patch} and \citet{Chandra}. At higher densities still, these differences are even larger and \htp\  electron donation becomes even more important. The contribution of \htp\  line absorption to the Rosseland mean opacity peaks at about $\log \rho=-7.5$ and T=3125 K, at which it is 15 \% of the opacity.

We give Rosseland mean opacities for a range of densities, temperatures and hydrogen mass fractions in table \ref{tab:ross_opa}.

\section{Stellar evolution calculations.}

We use the 1D CESAM stellar evolution code \citep{Morel} for our calculations. 
The CESAM code has been widely applied to Helio- and asteroseismology, but has also been used for studies low mass stars (0.6 \MO) \citep{Morel_LM}.

The basic input physics is identical to that described by  \citep{Morel}, except for the nuclear reaction rates, opacities and equations of state, which are as follows.
We use the recent NACRE \citep{NACRE1} compilation of nuclear reaction rates, which had been previously incorporated into the CESAM code and used for Solar evolution models \citep{Morel_NACRE}.
The recently improved and extended OPAL equation of state \citep{OPAL_EQS} which we refer to as OPAL02 is used for all temperatures and densities, from the core to the surface of the star. The most recent OPAL opacities \citet{OPAL_opa} which we refer to as OPAL96, are used for temperatures above 9000 K and the opacity data discussed in the previous section is used for T $<$ 9000 K. The equation of state discussed in section \ref{subsec:EoS} is used solely to calculate the number densities for use in the calculation of the low temperature opacities (T $<$ 9000 K). The OPAL02 equation of state provides the various thermodynamic quantities necessary for the calculation.
To investigate the effects of \htp\  on these models we calculate models using 2 different low temperature opacity functions. Firstly we perform a set of calculations using the \citet{Neale95} \htp\  partition function and secondly we repeat the calculation while completely neglecting \htp.

It should be noted that the that the OPAL96 opacities are limited to values of $\log R\leq$ 1, where $R=\rho/(T^3\times10^{-18})$. As the mass of a star decreases its density increases, such that for low metallicity stars of about 0.7~\MO\  and below, values of $\log R$ exceed 1 and can exceed 2 in a star of 0.4~\MO. In such circumstances we have extrapolated the OPAL96 opacity tables to higher values of $\log R$. An extrapolation to $\log R=2$ requires an extrapolation of an order of magnitude in density or an extrapolation of just over a factor of 2 in temperature. However the region in which we are forced to extrapolate the OPAL96 opacities corresponds to the convective regions of the stars. As convective, and not radiative, energy transport is dominant the extrapolation of the OPAL96 opacities is unimportant.
To test that this extrapolation is unimportant we perturbed the extrapolated OPAL96 opacity by multiplying by $\log R$ when $\log R\geq$1. We found that the calculations performed with and without this opacity perturbation were identical.

The minimum mass of star that we study is imposed by the boundaries of the OPAL02 equation of state tables.
These equation of state tables do not cover all the temperatures and densities within our model stars of mass less than about 0.15 \MO. Hence we limit our calculation to masses of 0.15 \MO\  and above.

A gray atmosphere is used for the surface boundary condition, the temperature stratification follows a T($\tau$) relationship based on Hopf's law. A thorough discussion of this and other model atmospheres used in CESAM can be found in \citet{Morel_atm}. In the zero metallicity model atmospheres of \citet{Saumon}, significant differences between the gray and non-gray ($\log(\rm g)=5$) model atmospheres do occur for \Teff\  $<$ 4500 K. Consequently for stars below this \Teff\  gray atmospheres give relatively poor boundary conditions. Non-gray atmospheres effectively increase the average atmospheric opacity relative to the gray case \citep{Saumon}. In this work the inadequacies of the gray atmosphere are relatively unimportant, because this is primarily a differential study in which we seek to show that the use of the \htp\  partition function of \citet{Neale95} causes \htp\  to have a greater effect on stellar evolution than previously thought. To enable us to calculate more realistic evolutionary models, we are in the process of calculating non-gray atmospheres for zero metallicity very low mass stars. 

The initial conditions for our calculation are those of a collapsing pre-Main Sequence star at the top of the Hayashi track, as described in \citet{Morel}. Briefly the pre-Main Sequence star is in quasi-static equilibrium and energy is released only through gravitational collapse. The star is fully convective so that entropy is constant throughout the star. The initial conditions are described by a contraction constant, c. For a 1 \MO\  star \citet{Morel} suggests a contraction constant of 0.02 L$_\odot$M$_\odot^{-1}$K$^{-1}$. In this work we use a constant value of 0.015  L$_\odot$M$_\odot^{-1}$K$^{-1}$, we find that the value of the contraction constant has only a small effect the position of the resultant star upon the Main Sequence. 

Our models are computed without mass loss and rotation. We follow \citet{Marigo} and use an initial hydrogen mass fraction of X$=$0.77. We set the metal mass fraction to be Z$=10^{-14}$, this is low enough to make the effect of metals negligible, whist avoiding numerical problems with extremely low and zero values of Z. This leaves a Helium mass fraction (Y) of marginally under 0.23. 

\section{Results and Discussion.}

\begin{figure}
\epsscale{0.9}
\plotone{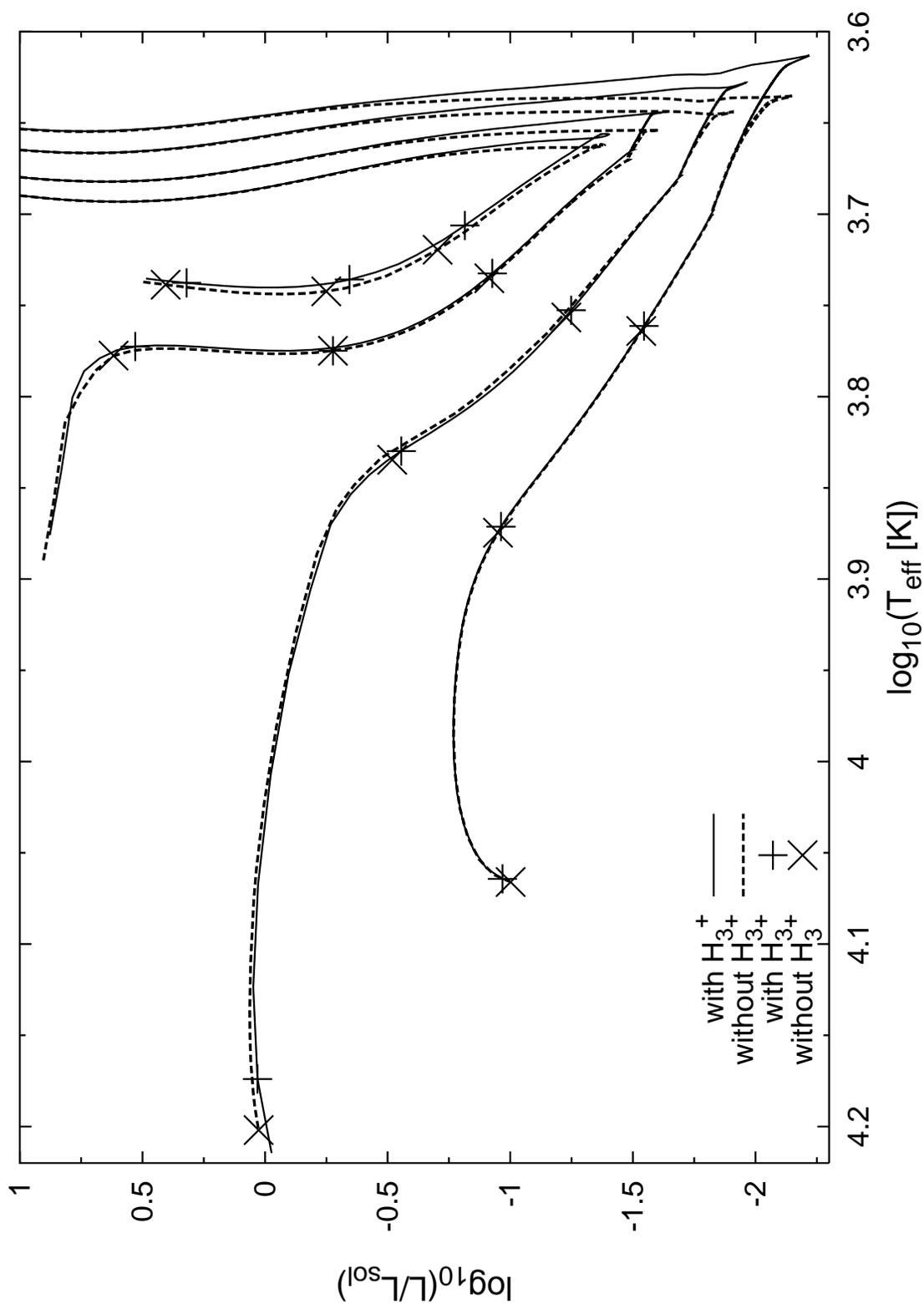}
\caption{The evolutionary tracks of 0.4, 0.3, 0.2 and 0.15 solar mass zero metallicity stars calculated by both neglecting and including the effect of \htp\  upon opacity. Symbols are placed at values of the hydrogen mass fraction of 10$^{-2}$, 10$^{-3}$ and 10$^{-4}$ at the center of the star.}
\label{fig:HR.to.0.4.ps}
\end{figure}

Figure \ref{fig:HR.to.0.4.ps} shows the evolution paths of zero metallicity model stars between 0.4 and 0.15 \MO\  calculated with and without the effects of \htp. Symbols are placed on the curves when the core hydrogen mass fractions are 10$^{-2}$, 10$^{-3}$ and 10$^{-4}$. Initially the pre-Main Sequence stars of 0.4 \MO\  and less, with and without \htp\  contract along identical paths, the density within the atmosphere at this stage is too low for \htp\  to have any effect. As the star becomes more dense and before the stars reach the Main Sequences the paths separate as \htp\  starts to become the dominant positive ion within the atmosphere. At zero age main sequence the star with \htp\  has a lower \Teff\  and luminosity, than the star in which \htp\  is not accounted for. This trend continues throughout the Main Sequence lifetime of the stars. For stars with higher masses of 0.5 and 0.45 \MO\  we find that \htp\  affects the late pre-Main Sequence and early Main Sequence evolution, but that the evolutionary tracks on the Main Sequence converge as the star becomes hotter in the late main sequence, and \htp\  is destroyed. Our evolution tracks are in agreement with the zero metallicity calculations of \citet{Marigo} which cover masses greater than 0.7 \MO\  and also with \citet{Siess} which cover masses greater than 0.8 \MO.

\begin{figure}
\epsscale{0.9}
\plotone{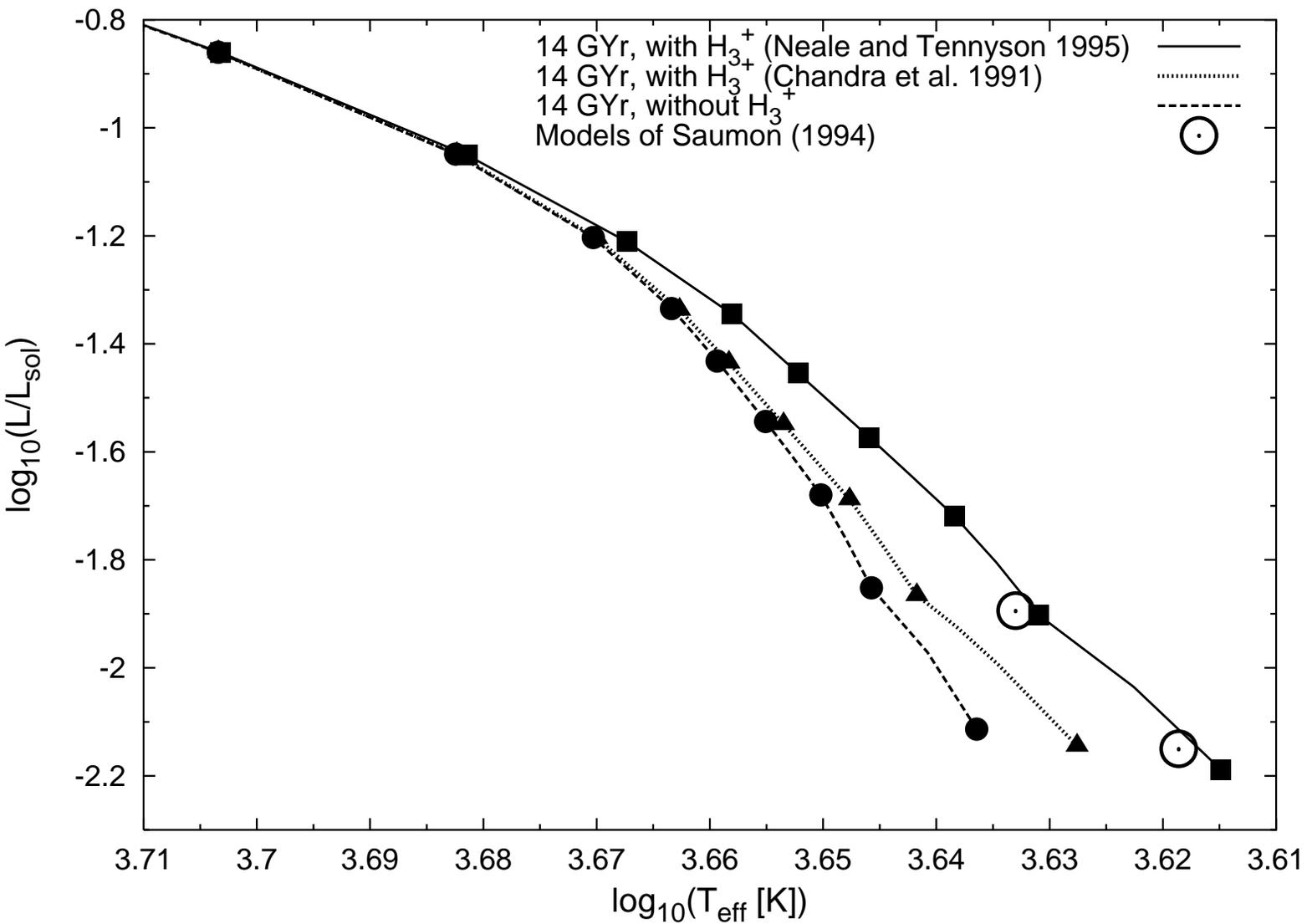}
\caption{The 14 GYr isochrone for zero metallicity stars of mass 0.55 to 0.15 \MO. Symbols are placed between masses of 0.55 and 0.15 \MO, in steps of 0.05 \MO. Also shown are the 0.2 \MO\  and 0.15 \MO\  zero metallicity Main Sequence models of \citet{Saumon}.}
\label{fig:isoc}
\end{figure}

This tendency for model stars of less than 0.5 \MO\  with  \htp\  to have lower \Teff\  and luminosity than their counterparts without \htp\  is shown again in figure \ref{fig:isoc}. This figure shows the 14 GYr isochrone for model stars calculated with and without the effects of \htp, with symbols placed at intervals of 0.05 \MO. The three curves represent models calculated using the \citet{Neale95} \htp\  partition function, the \citet{Chandra} \htp\  partition function and also neglecting \htp. 
For 14 GYr old stars of 0.15 \MO, the difference in \Teff\  between models calculated with the \citet{Neale95} partition function and those in which \htp\  is neglected is over 200 K and the difference in luminosity is about 18 \%. 
The models calculated with the \citet{Neale95} partition function show that \htp\  has a significantly greater effect, which extends to higher masses, than that predicted by models calculated with the \citet{Chandra} partition function. 
This illustrates the importance of using the \citet{Neale95} partition function in low mass evolution models. The luminosity and effective temperature of the 0.2 \MO\  and 0.15 \MO\  zero metallicity Main Sequence model stars of \citet{Saumon} are also plotted in figure \ref{fig:isoc}. As \citet{Saumon} used the \citet{Chandra} partition function, their results would have been expected to lie upon the isochrone calculated with the \citet{Chandra} partition function. However the models of \citet{Saumon} lie closer to the isochrone calculated with the \citet{Neale95} partition function. \citet{Saumon} used non-gray atmospheres whereas in this work we have used gray atmospheres, so the reason for this difference is likely to be due to the break down of the gray approximation in very low mass stars. \citet{Saumon} commented that using non-gray models effectively increases the atmospheric opacity over gray model atmospheres. Hence non-gray stars have a lower \Teff\  than their gray counterparts.

It is clear from figure \ref{fig:isoc} that the strength of the effect of \htp\  increases with decreasing mass. As long as H$^-$ remains the dominant source of opacity the strength of the effect of \htp\  will continue to increase with decreasing mass. The effect of \htp\  within the atmosphere will diminish when collision induced absorption becomes the dominant source of opacity, at about T$_{\rm eff}<3000$ K. However in a shell below the photospheres of such stars hotter denser gas exists in which H$^-$ remains the dominant source of opacity and \htp\  will again affect opacity. 

\begin{figure}
\epsscale{0.9}
\plotone{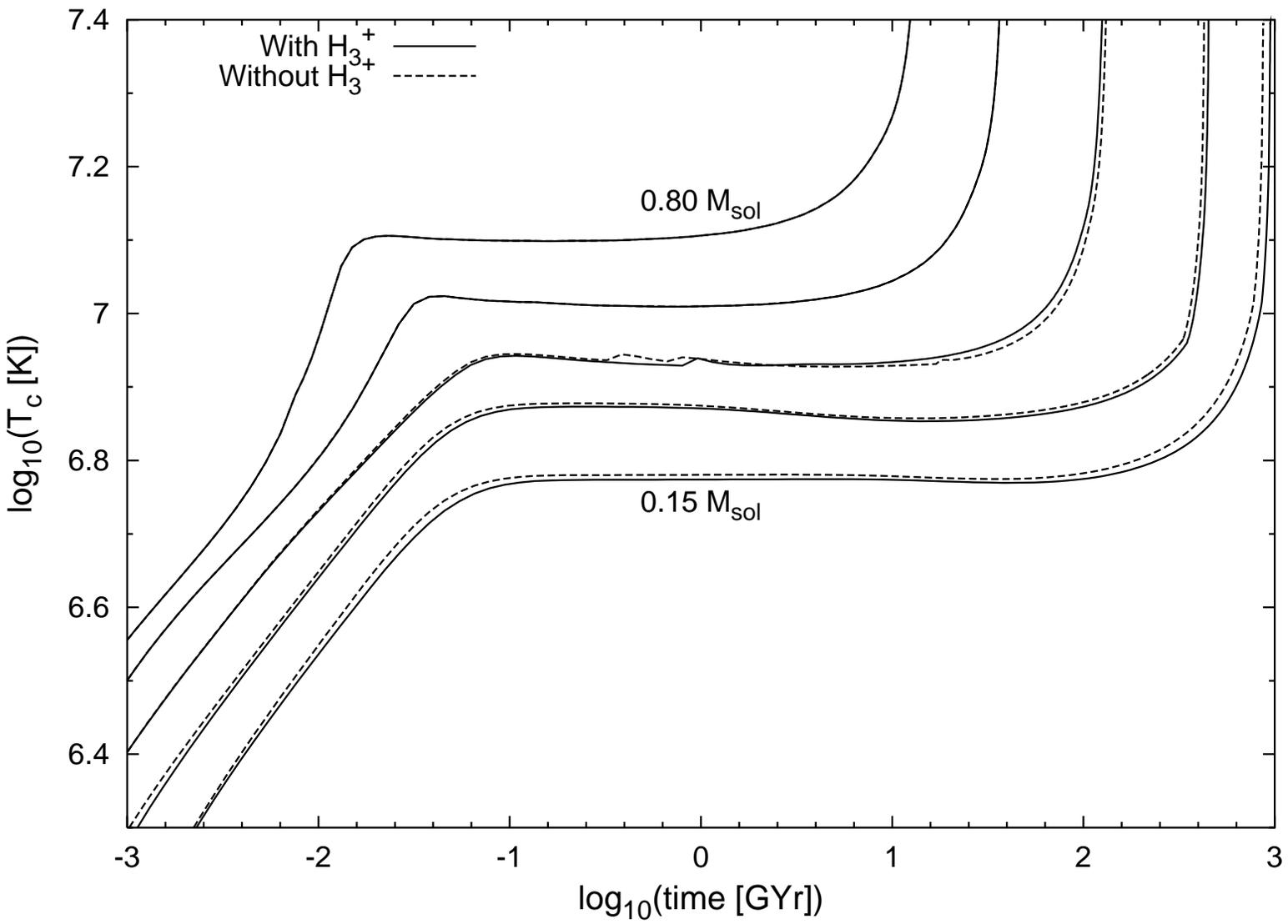}
\caption{The central temperature as a function of time for stars of masses 0.8, 0.6, 0.4, 0.25 and 0.15 \MO\, curves are shown for including and neglecting \htp.}
\label{fig:Tc}
\end{figure}

Figure \ref{fig:Tc} shows the evolution of the central temperature for model stars of mass 0.8, 0.6, 0.4, 0.25 and 0.15 \MO, calculated with and with out the effects of \htp. Again it is seen that \htp\  has no effect on the the central temperatures of model stars of mass 0.6 \MO\  and above. For the model stars of 0.25 and 0.15 \MO\  the increased atmospheric opacity due to \htp\  acts to reduce the central temperature. This in turn reduces the nuclear reaction rates and so increases the hydrogen burning lifetime of the model star. \htp\  also acts to reduce the pre-Main Sequence and early Main Sequence central temperatures of the 0.4 \MO\  model. However for most of the Main Sequence the central temperature of the 0.4 \MO\  model star which neglects \htp\  is cooler than the 0.4 \MO\  model which accounts for \htp. This results in a longer hydrogen burning lifetime for the model which neglects \htp. The reason for this is that the  0.4 \MO\  model star without \htp\  is fully convective at zero age Main Sequence. Whereas the 0.4 \MO\  model star which accounts for \htp\  starts zero age Main Sequence with a radiative interior, but the model star which neglects \htp\  starts zero age main sequences with a fully convective interior. Convective mixing in a fully convective star makes the hydrogen throughout the star available for fusion in the core. Conversely in a star with a radiative interior there is only very limited mixing of material, so that only the hydrogen in the core is available for fusion. Hence the convective model star without \htp\  has a longer lifetime than the radiative model star with \htp, even though the former is more luminous. It appears that \htp\  within the atmosphere, can have an effect upon the structure and energy transport mechanisms within these low mass stars that have interior temperature gradients close to the adiabatic temperature gradient. So slight changes in the temperature gradient brought about by increased atmospheric opacity can change the main energy transport method. All the stars of mass less than 0.4 \MO\  have a fully convective interior at zero age main sequence, all of the stars develop a radiative interior at some point later in their main sequence evolution. After developing a radiative interior the rate of evolution of the star along the main sequence increases dramatically.

\section{Conclusion.}

We have calculated new low temperature zero metallicity Rosseland mean opacities using the latest collision induced absorption opacity data and with the latest \htp\  partition function and line opacity data. The results of this calculation confirm that at high densities and intermediate temperatures \htp\  becomes the dominant positive ion.
Furthermore using the \htp\  \citet{Neale95} partition function instead of an older partition function, causes the temperature and density ranges at which \htp\  forms to expand as well as increasing its abundance by up to an order of magnitude. This in turn increases the abundance of electrons and hence H$^-$ and other negative ions, thereby affecting the opacity of the gas by a greater degree than was previously thought. The \htp\  line opacity at certain temperatures and densities can contribute up to 15 \% of the Rosseland mean opacity. \htp\  lines may well be observable within cool Population III stars and all models of such stars should account for direct absorption by \htp.

The new Rosseland mean opacities have been included in evolution models of very low mass zero metallicity stars, enabling us to investigate the effect of \htp\  on the evolution of these stars. For late pre-Main Sequence and early Main Sequence stars of mass $<$ 0.55 \MO\  we find that \htp\  can cause the atmosphere to cool and the star to drop slightly in luminosity. For stars of 0.45 \MO\  and below this effect becomes strong enough to affect the Main Sequence structure and lifetime of the star.

The results presented here show that as the mass of a zero metallicity star is reduced the effect of \htp\  upon its evolution, structure, luminosity and temperature increase. As the \citet{Neale95} partition function is much larger than the partition function of \citet{Chandra}, used by \citet{Saumon}, its corresponding effect on opacity is larger. Hence if the effect of \htp\  continues to grow in objects of below 0.15 \MO, then \htp\  will effect the evolution of zero metallicity very low mass stars and brown dwarfs too a greater extent than previously believed.

\acknowledgments

Our anonymous referee is thanked for their constructive criticism which has helped to improve this paper. We thank Dr. P. Morel for use of his stellar evolution code CESAM and Dr. A. Borysow for making the results of her CIA calculations publicly available. We thank the UK Particle Physics and Astronomy Research Council (PPARC) for post-doctoral funding for GJH and funding for the computer hardware upon which our calculations have been performed. The calculations reported here were, in part, carried out using the Miracle 24-processor Origin 2000 supercomputer at the HiPerSPACE computing center, UCL, which is part funded by PPARC.

\begin{table}
\caption{The species used in the calculation, the source of their rotational-vibrational partition functions is quoted where applicable.}

\begin{tabular}{lcl}
\hline
Species                    & T range (K) & Reference \\ \hline
  H$_2$                    & 1000--9000  & \citet{Sauval} \\
  H                        &  -          & - \\
  H$^-$                    &  -          & - \\
  H$^+$                    &  -          & - \\
  H$_2^+$                  & 1500--18000 & \citet{Stancil} \\
  H$_2^-$                  & 1000--9000  & \citet{Sauval} \\ 
  \htp                     & 500--8000   & \citet{Neale95} \\
  He                       &        -    & -  \\
  He$^+$                   &       -     & -  \\
  HeH$^+$                  & 1000--9000  & \citet{Sauval} \\
  e$^-$                    &       -     & -  \\ \hline
\end{tabular}

\label{tab:partfunc}

\end{table}

\begin{table}
\caption{The sources of continuous monochromatic opacity used in the calculation, with temperature and wavenumber limits if applicable.}

\begin{tabular}{lccl}
\hline
&
\multicolumn{2}{c}{Limits} & \\ \cline{2-3}

Opacity                     & T (K)       & wavenumber (\cm)   & Reference \\ \hline
  H$_2$-He CIA              & 1000 -- 7000   & 25 -- 20~000 & \citet{CIA_H2He} \\
  H$_2$-H$_2$ CIA           & 1000 -- 7000   &  20 -- 20~000 & \citet{CIA_H2H2} \\
  H-He CIA                  & 1500 -- 10~000 &  50 -- 10~000 & \citet{CIA_HHe} \\
  
  H$^-$ bound-free          & --             &  $<$ 80~000  & Fits of \citet{Hm} \\
  H$^-$ free-free           & 1400 -- 10~000 &  $<$ 55~000  & Fits of \citet{Hm} \\

  He$^-$ free-free          & 660 -- 22~000 & 1400 -- 10~000 & \citet{Bell82} \\

  He bound-free             & --         &   --  & \citet{Kurucz1970} Method \\
  He free-free              & --         &   --  & \citet{Kurucz1970} Method \\

  He$^+$ bound-free         &   --      &  --   & \citet{Kurucz1970} Method  \\
  He$^+$  free-free         &   --      &  --   & \citet{Kurucz1970} Method \\

  H bound-free              &   --       &  --   & Method of \citet{Gray} \\
  H free-free               &   --       &  --  & Method of \citet{Gray} \\ 
 & & & using Gaunt factors from \\
 & & & \citet{Karzas} \\

  H$_2^-$ free-free         & 658 -- 28~500 & $>$ 1400 -- 10~000   & \citet{Bell80} \\
  H$_2^+$ b-f, f-f & 4500 -- 10~000 & 2000 -- 46~000 & \citet{Lebedev} \\

  H$_2$ Rayleigh scat. & --        & $>$ 9.76 & \citet{Kurucz1970} method. \\
  H  Rayleigh scat.    & --          & $<$ 82~000 & \citet{Kurucz1970} method. \\
  He Rayleigh scat.    & --  &    $>$ 17.2 & \citet{Kurucz1970} method.  \\

  e$^-$ Thomson scat.  & --         & -- & Method of \citet{Gray} \\
\htp\ Line opacity          & --         & $<$ 15~000 & \citet{Neale96} \\ 
H I Line opacity            & --         & --         & Subroutine from ATLAS12 \\
                            &            &            & \citet{Kurucz1993} \\  \hline
\end{tabular}

\label{tab:opasour}

\end{table}

\begin{table}
\caption{Rosseland mean opacities ($\kappa_R$) for a range of hydrogen mass fractions (0 $\le$ X $\le$ 1), densities ($10^{-14}\le \log_{10}(\rho\  [{\rm g~cm^{-3}}]) \le 10^{-2}$) and temperatures (1000 $\le$ T(K) $\le$ 9000 ). The complete version of this table is in the electronic edition of the Journal.  The printed edition contains only a sample. }

\begin{tabular}{llll}
\hline
X & $\log_{10}(\rho~[{\rm g~cm^{-3}]})$ & T (K) & $\log_{10}(\kappa_R~[{\rm cm}^2 \  { \rm g}^{-1}])$  \\ 
\hline
0.70E+00 & -5.50E+00 & 2.00E+03 & -4.158E+00 \\
0.70E+00 & -5.50E+00 & 2.25E+03 & -4.181E+00 \\
0.70E+00 & -5.50E+00 & 2.50E+03 & -4.189E+00 \\
0.70E+00 & -5.50E+00 & 2.75E+03 & -4.179E+00 \\
0.70E+00 & -5.50E+00 & 3.00E+03 & -4.095E+00 \\
0.70E+00 & -5.50E+00 & 3.25E+03 & -3.802E+00 \\
0.70E+00 & -5.50E+00 & 3.50E+03 & -3.374E+00 \\
0.70E+00 & -5.50E+00 & 3.75E+03 & -2.973E+00 \\
0.70E+00 & -5.50E+00 & 4.00E+03 & -2.593E+00 \\
0.70E+00 & -5.50E+00 & 4.25E+03 & -2.157E+00 \\
0.70E+00 & -5.50E+00 & 4.50E+03 & -1.738E+00 \\
0.70E+00 & -5.50E+00 & 4.75E+03 & -1.362E+00 \\
0.70E+00 & -5.50E+00 & 5.00E+03 & -1.031E+00 \\
0.70E+00 & -5.50E+00 & 5.25E+03 & -7.399E-01 \\
0.70E+00 & -5.50E+00 & 5.50E+03 & -4.811E-01 \\
0.70E+00 & -5.50E+00 & 5.75E+03 & -2.486E-01 \\
0.70E+00 & -5.50E+00 & 6.00E+03 & -3.780E-02 \\
0.70E+00 & -5.50E+00 & 6.25E+03 &  1.552E-01 \\
0.70E+00 & -5.50E+00 & 6.50E+03 &  3.335E-01 \\
0.70E+00 & -5.50E+00 & 6.75E+03 &  4.994E-01 \\
0.70E+00 & -5.50E+00 & 7.00E+03 &  6.551E-01 \\
 \hline
\end{tabular}

\label{tab:ross_opa}

\end{table}

\end{document}